\documentclass[runningheads]{llncs}
\usepackage{graphicx}
\usepackage{caption}
\usepackage{booktabs}
\usepackage{ amssymb }
\usepackage{bbold}
\usepackage{gensymb}
\usepackage{multirow}
\usepackage{multicol}
\usepackage{marvosym}
\usepackage{xcolor}
\usepackage[hidelinks,colorlinks,citecolor=gray,linkcolor=gray]{hyperref}
\usepackage{amsmath}
\usepackage{enumitem}

\begin{document}
\title{FairPrune: Achieving Fairness Through Pruning for Dermatological Disease Diagnosis}
\titlerunning{FairPrune: Achieving Fairness Through Pruning}
%

\author{Yawen Wu\inst{1*}
\and
Dewen Zeng\inst{2*}
\and
Xiaowei Xu\inst{3}
\and
Yiyu Shi\inst{2}$^{(\textrm{\Letter})}$
\and
Jingtong Hu\inst{1}$^{(\textrm{\Letter})}$
}

\institute{University of Pittsburgh, Pittsburgh, PA, USA \\\email{\{yawen.wu, jthu\}@pitt.edu}\and
University of Notre Dame, Notre Dame, IN, USA 
\\\email{\{dzeng2, yshi4\}@nd.edu} \and 
Guangdong Provincial People's Hospital, Guangzhou, China \\ \email{xxu8@nd.edu} \\
* Equal contributions. Listing order determined by coin flipping.
}
\authorrunning{Y. Wu, D. Zeng et al.}
\maketitle              
\begin{abstract}
Many works have shown that deep learning-based medical image classification models can exhibit bias toward certain demographic attributes like race, gender, and age. Existing bias mitigation methods primarily focus on learning debiased models, which may not necessarily guarantee all sensitive information can be removed and usually comes with considerable accuracy degradation on both privileged and unprivileged groups. 
To tackle this issue, 
we propose a method, FairPrune, that achieves fairness by pruning.
Conventionally, pruning is used to reduce the model size for efficient inference. 
However, we show that pruning can also be a powerful tool to achieve fairness.
Our observation is that during pruning, each parameter in the model has different importance for different groups' accuracy.
By pruning the parameters based on this importance difference, 
we can reduce the accuracy difference between the privileged group and the unprivileged group to improve fairness without a large accuracy drop. 
To this end, we use the second derivative of the parameters of a pre-trained model to quantify the importance of each parameter with respect to the model accuracy for each group. Experiments on two skin lesion diagnosis datasets over multiple sensitive attributes demonstrate that our method can greatly improve fairness while keeping the average accuracy of both groups as high as possible.

\end{abstract}

\section{Introduction}

In AI-assisted medical image analysis, deep neural networks (DNNs) tend to capture relevant statistical information such as colors and textures from the training data.
This data-driven paradigm can help the network learn task-specific features for high accuracy on the target task.
However, to maximize the accuracy,
the network may use the information present in some data but not in other data,
and thus show discrimination towards certain demographics (i.e. skin tone or gender).
For example, \cite{du2021fairness,xu2020investigating} demonstrate that the network learned on the CelebA dataset \cite{liu2015faceattributes} turns to perform better on the female group when the task is predicting facial attributes such as wavy hair and smiling.
Dermatological disease classification networks trained on two public dermatology datasets Fitzpatrick-17k and ISIC 2018 Challenge have been reported to be biased across different skin tones \cite{groh2021evaluating,kinyanjui2020fairness}. 
However, no solution is proposed to mitigate the bias in these works.
An X-ray computer-aided diagnosis (CAD) system was found to exhibit disparities across genders \cite{larrazabal2020gender}.
Once these biased models are deployed in the real-world system, they could be harmful to both individuals and society.
For example, AI algorithms could misdiagnose people with different demographic groups, leading to increased health care disparities.
This leads to a variety of research techniques that aim to alleviate the biased in DNNs.

One of the most widely used bias mitigation methods is adversarial training \cite{kim2019learning,wang2019balanced,elazar2018adversarial,alvi2018turning,zhang2018mitigating}.
Normally, an adversarial network is added to the tail of an encoder or a classifier to predict the protected attributes and form a minimax game: maximize the network's ability to predict the class while minimizing the adversarial network's ability to predict the protected attributes.
In this way, the model will be able to learn fair features that are irrelevant to the protected attributes.
However, as mentioned in \cite{wang2020towards}, the main drawback of this method is that even if the target protected attribute has been removed, the combination of other features may still be a proxy of this protected attribute.
In addition, forcing the model to ignore the protected attribute relevant features may harm its classification accuracy as those features may contain important information for the final prediction \cite{wang2020towards}.
Fairness through explanation is another bias mitigation technique \cite{rieger2020interpretations,hendricks2018women,singh2020don}, this technique requires fine-grained feature-level annotation as the domain knowledge to train the model to only focus on bias-unrelated features in the original input. 
Such suppression of sensitive information can also potentially remove useful information and thus greatly degrade the classification performance.
Therefore, to achieve fairness, these state-of-the-art (SOTA) methods usually need to sacrifice considerable accuracy for both groups.

To avoid this, we introduce FairPrune, a technique to achieve fairness via pruning.
Conventionally, pruning is used to reduce the model size for efficient inference. 
However, one interesting thing we found is that pruning can be a powerful tool for fairness.
Our work is motivated by the observation that during pruning, the parameter that is important for one demographic group may be unimportant for another.
By controlling the parameters to prune, we can reduce the accuracy difference between the privileged group and the unprivileged group to improve fairness while keeping their overall accuracy as high as possible.
To this end, we utilize the saliency of each parameter (computed based on the second derivative \cite{lecun1989optimal}) to quantify its importance regarding the model accuracy. 
Specifically, we compute the saliencies of all parameters for each demographic group, which will then be used to prune the parameters that show large importance differences for these two groups to mitigate biases.
Besides, the trade-off between fairness improvement and accuracy drop can be adjusted to satisfy different user requirements. 
As a byproduct, FairPrune can also reduce the network size for efficient deployment.
We evaluate FairPrune on two skin lesion analysis datasets over two sensitive attributes and show improved fairness with a lower accuracy drop over SOTA methods for fairness.

\section{Related Work}

Existing bias mitigation methods can be generally categorized into three groups: pre-processing, in-processing, and post-processing.

For pre-processing methods, a straightforward solution is to remove the sensitive information from the training data. One could also assign different weights to different data samples to suppress the sensitive information during training \cite{kamiran2012data}.
These pre-processing techniques are not suitable for dermatological data because 
the sensitive information exists in the target (i.e., diseased area), which is necessary for diagnosis purposes and cannot be removed.

In-processing bias mitigation methods usually involve modifying the training loss function to regularize the model for fairness.
For example, \cite{liu2019incorporating} added a task-specific prior to implicitly regularizing the model not to pay attention to the sensitive related information for its prediction. 
Adversarial training \cite{zhang2018mitigating,tzeng2015simultaneous,wang2019balanced,kim2019learning} achieve fairness by removing sensitive information with an adversarial learner.
However, these methods can not explicitly protect the unprivileged group when enforcing the fairness constraints, and the accuracy of both groups will drop.
Instead, our method can protect the accuracy of one group when achieving fairness goals, resulting in high overall accuracy and fairness simultaneously.

As for post-processing techniques, 
during the inference-time, calibration is performed by taking the model's prediction and the sensitive attribute as input \cite{hardt2016equality,zhao2017men,du2020fairness}. 
The goal is to enforce the prediction distribution to match the training distribution or a specific fairness metric.
While these methods show effectiveness for bias mitigation, they need access to the sensitive attribute for inference, which may not be available for disease diagnosis when taking images as input.
Different from these methods, our method modifies the pre-trained model only based on the training set. During inference, the sensitive attribute is not needed.

\section{Method}
\subsection{Problem Definition}
Given a dataset $D=\{x_i,y_i,c_i\}$, $i \in 1,...,N$ where $x_i$ is the input image, $y_i$ is the class label, $c_i$ is the sensitive attribute (e.g., skin tone, gender, age), and a pre-trained classification model $f_\theta(\cdot)$ with parameters $\theta$ that maps the input $x_i$ to the final prediction $\hat{y_i}=f_\theta(x_i)$.
Our goal is to reduce the discrimination in $f_\theta(\cdot)$ with respect to the sensitive attribute $c$ by only modifying some of the parameters in $f_\theta(\cdot)$ without further finetuning. In this paper, we only consider the binary sensitive attribute (i.e., $c_i \in \{0,1\}$). $c_i=0$ represents unprivileged samples (the model shows discrimination against), while $c_i=1$ represents privileged samples.

\subsection{Pruning for Fairness}

\noindent
\textbf{Saliency reflects the accuracy drop after pruning.}
Given the pre-trained model $f_\theta$ and its objective function $E$,
the change of the objective function after pruning parameters $\Theta$ can be approximated by a Taylor series \cite{lecun1989optimal}:
\setlength{\abovedisplayskip}{0pt}
\setlength{\belowdisplayskip}{0pt}
\setlength{\abovedisplayshortskip}{0pt}
\setlength{\belowdisplayshortskip}{0pt}
\begin{equation}\label{equ:objective_delta}
\begin{split}
\Delta E & = E(D | \Theta = 0) - E(D) \\
& = -\sum_i g_i \theta_i + \frac{1}{2}\sum_i h_{ii} \theta_i^2 + \frac{1}{2} \sum_{i\neq j}h_{ij}\theta_i \theta_j + O(||\Theta||^3) = \frac{1}{2} \sum_i h_{ii} \theta_i^2.
\end{split}
\end{equation}
where $g_i = \frac{\partial E}{\partial \theta_i}$ is the gradient of $E$ with respect to $\theta_i$, which is close to 0 because we assume the pre-trained model has converged and the objective function is at its local minimum. $h_{ii} = \frac{\partial^2 E}{\partial^2 \theta_i}$ is the element at row $i$ and column $i$ of second derivative Hessian matrix $\textbf{H}$.
The approximation assumes that $\Delta E$ caused by pruning several parameters is the sum of $\Delta E$ caused by pruning each parameter individually, so the third term is neglected.
$\frac{1}{2} h_{ii} \theta_i^2$ is called the saliency of $\theta_i$, which represents the increase of error after pruning this parameter.

\begin{figure}[!t]
	\centering
	\includegraphics[width=0.95\columnwidth]{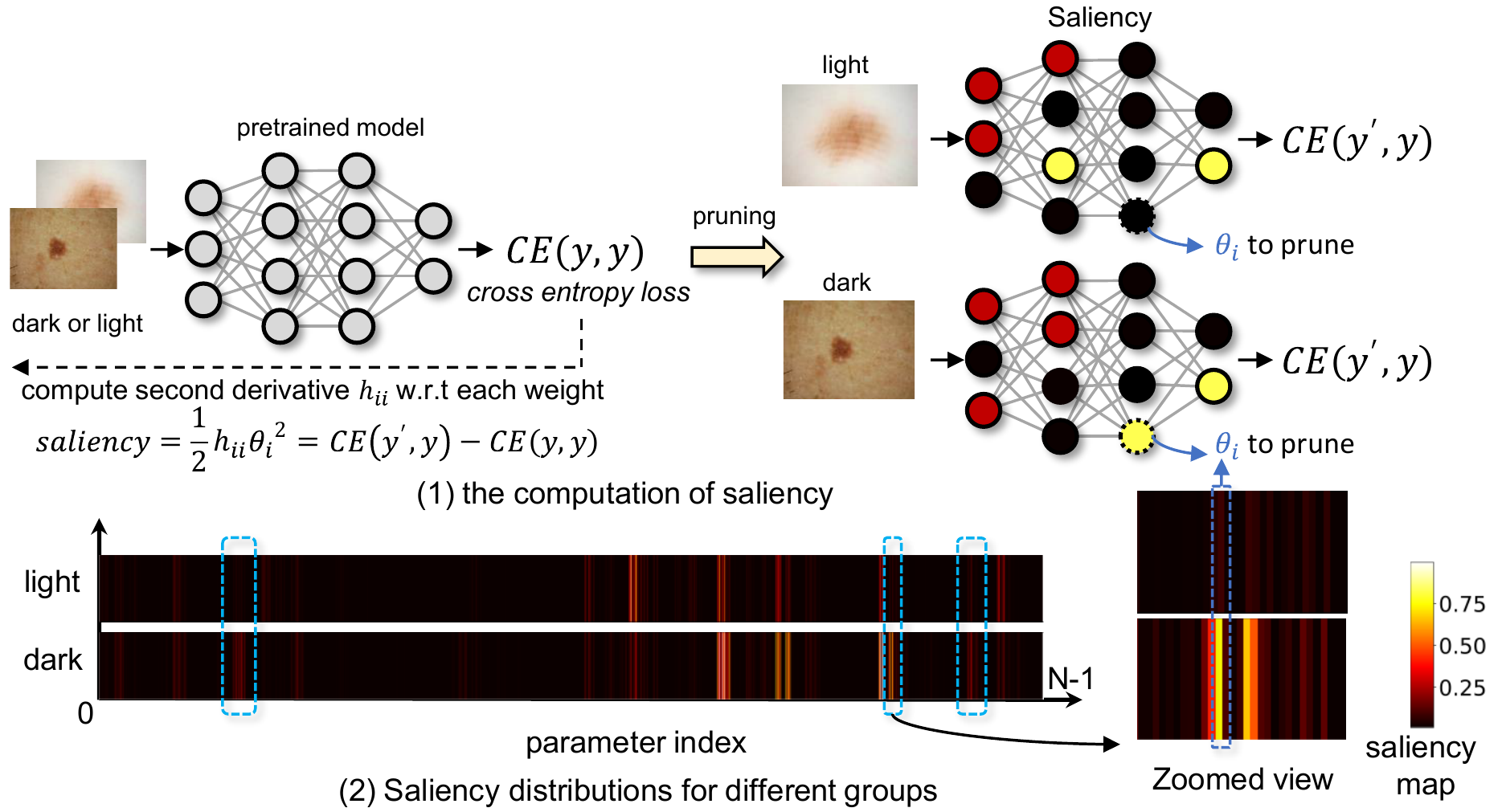}
	\caption{Illustration of (1) saliency computation and (2) the parameters' saliency distribution for dark and light skin tone groups on the Fitzpatrick-17k dataset. The x-axis is the index of parameters in the first layer of the VGG-11, the color represents the normalized saliency for each demographic group. The blue boxes highlight those parameters that have relatively low saliency for the light (unprivileged group) but high saliency for the dark (privileged group). }
	\label{fig:example_saliency}
\end{figure}

\noindent
\textbf{Using saliency to achieve group fairness.}
Our idea is based on our empirical observation that in a pre-trained network, the importance (saliency) of some parameters can be totally different for different demographic groups. 
That is, some parameters may have a small saliency for one group but a large saliency for another group.
For example, Fig. \ref{fig:example_saliency} shows a dermatological disease diagnosis model pre-trained by vanilla training and is biased for patients with light skin tone, which is the unprivileged group.
In the coordinate, we show the saliency distribution of the parameters in the first layer of VGG-11 for both dark and light groups.
The x-axis represents parameter indices, where one coordinate shows the same parameter for both the light and the dark groups.
The color denotes normalized saliency for both groups.
It can be seen that there exist differences between the distribution of saliency for these two groups.
In the blue boxes, we highlight those parameters with low saliency for the light group but high saliency for the dark group.
By pruning these parameters, the accuracy difference of the privileged group (dark) and the unprivileged group (light) can be reduced. In this way, the model is pruned to be fair for the unprivileged group.

To identify those parameters to prune, we propose a fairness-aware saliency computation method to identify the parameters which are unimportant for the unprivileged group but important for the privileged group. The importance of each parameter for fairness is quantified by integrating the saliency of the unprivileged group and the saliency of the privileged group weighted by a negative scalar.
To be specific, we formulate a multi-objective optimization problem:
\begin{equation}\label{equ:multi_objective}
\min \Delta E_{c=0}(\Theta), \quad \max \Delta E_{c=1}(\Theta).
\end{equation}
where $\Delta E_{c=0}(\Theta)$ and $\Delta E_{c=0}(\Theta)$ are the error changes of unprivileged group $c=0$ and privileged group $c=1$ after pruning parameters $\Theta$, respectively.

To solve this problem, Eq.(\ref{equ:multi_objective}) can be transformed into a single objective as:
\begin{equation}
\min_{\Theta} J = \Delta E_{c=0}(\Theta) - \beta \Delta E_{c=1}(\Theta) = \sum_{i} s_i,
\end{equation}
\begin{equation}\label{equ:combined_score}
    s_i = (\frac{1}{2}h_{ii}^0 \theta_i^2 ) - \beta \cdot (\frac{1}{2}h_{ii}^1 \theta_i^2 ) = \frac{1}{2}\theta_m^2 (h_{ii}^0 - \beta \cdot  h_{ii}^1).
\end{equation}
where $\beta$ is a hyper-parameter controlling the trade-off between minimizing $\Delta E_{c=0}$ and maximizing $\Delta E_{c=1}$. 
$\theta_i$ is the parameter of the model and can be treated as a constant. 
$s_i$ is the saliency of parameter $\theta_i$ for achieving the fairness objective, where a smaller value represents a larger benefit when we prune it since a smaller $s_i$ contributes to minimizing the objective $J$.
$h_{ii}^c, c\in\{0,1\}$ is the Hessian element of $\theta_i$ for demographic group $c$, which is the juncture to inject the information of each group into the pruning process to achieve fairness.

\subsection{Pruning Recipe}
FairPrune prunes a pre-trained model by the following iterative steps.
\begin{enumerate}[topsep=0pt,itemsep=-1ex,partopsep=1ex,parsep=1ex]
    \item Sample mini-batches $\{ B^0 \}$ and $\{ B^1 \}$ from the unprivileged group and privileged group, respectively.
    \item For each pair of mini-batches $(B^0, B^1)$, compute the second derivatives $h_{ii}$ for each parameter, and compute parameter salience $s_i$.
    \item After several mini-batches, average the saliency of each parameter over the mini-batches. Remove $p\%$ parameters with the smallest saliency. 
    \item Repeat steps 1-3 until the target fairness metric is achieved.
\end{enumerate}

\section{Experiments and Results}

\textbf{Fairness Metrics.}
We use multi-class equalized opportunity (Eopp) and equalized Odds (Eodd) \cite{hardt2016equality} to evaluate the fairness of the model. 
The Eopp0 is the True Negative Rate difference between two groups, the Eopp1 is the True Positive Rate difference between two groups, while the Eodd is the summation of the True Positive Rate difference and False Positive Rate difference.
Suppose $TP_k^c$, $FN_k^c$, $TN_k^c$, and $FP_k^c$ are the True Positive, False Negative, True Negative, and False Positive of class $k$ and group $c$, then the True Positive Rate, True Negative Rate, and False Positive Rate of class $k$ and group $c$ can be computed by $TPR_k^c=\frac{TP_k^c}{TP_k^c+FN_k^c}$, $TNR_k^c=\frac{TN_k^c}{TN_k^c+FP_k^c}$, and $FPR_k^c=\frac{FP_k^c}{TN_k^c+FP_k^c}$, respectively.
The EOpp and Eodd can be computed by the following equations:
\begin{equation}
    EOpp0 = \sum_{k=1}^{K}|TNR_k^1-TNR_k^0|, \quad EOpp1 = \sum_{k=1}^{K}|TPR_k^1 - TPR_k^0|.
\end{equation}
\begin{equation}
    EOdd = \sum_{k=1}^{K}|TPR_k^1 - TPR_k^0 + FPR_k^1 - FPR_k^0|.
\end{equation}

\noindent
\textbf{Dataset and Preprocessing.}
The proposed methods are evaluated on two dermatology datasets for disease classification, including the Fitzpatrick-17k \cite{groh2021evaluating} and ISIC 2019 challenge \cite{combalia2019bcn20000,tschandl2018ham10000} datasets.
The Fitzpatrick-17k contains 16,577 images in 114 skin conditions.
The skin tones are categorized in six levels from 1 to 6, where a smaller value represents lighter skin and a larger value means darker skin. We categorize the skin tones into two groups, where types 1 to 3 are light skin and types 4 to 6 are dark skin.
The ISIC 2019 dataset contains 25,331 dermoscopic images in 9 diagnostic categories.
Since skin tones are not provided by this dataset, we use the gender label to group the data as female and male.

\noindent
\textbf{Pre-training Details.}
We use VGG-11 as the backbone. On both datasets, we resize all the images to 128$\times$128.
Data augmentation includes random horizontal flipping, vertical flipping, rotation, scaling, and autoaugment \cite{cubuk2018autoaugment}.
We randomly split the dataset into training (60\%), validation (20\%), and test (20\%) partitions.
The model is pre-trained for 200 epochs with the Adam optimizer.
The batch size is 256 and the learning rate is 1e-4 decayed by a factor of 10 at epoch 160.

\noindent
\textbf{Pruning Details.}
On the Fitzpatrick-17k dataset, the batch size for computing the saliency is 2. The saliency of each parameter is averaged every 500 mini-batches to prune $5\%$ of the parameters in each pruning iteration. 
On the ISIC 2019 dataset, the batch size for computing the saliency is 64. The saliency of each parameter is averaged every 200 mini-batches to prune $10\%$ of the parameters.
Grid search is used to find the best hyper-parameter $\beta$ and pruning ratio with the optimal accuracy and fairness tradeoff on the validation set for each dataset.
The pre-training and pruning are performed on one Nvidia V100 GPU.

\noindent
\textbf{Baselines.}
We compare FairPrune with multiple baselines.
\textit{Vanilla} is standard training without fairness constraints.
\textit{AdvConf} and \textit{AdvRev} are two adversarial training based de-biasing methods.
\textit{AdvConf} \cite{alvi2018turning,tzeng2015simultaneous} 
employs a uniform confusion loss to minimize the classifier's ability for predicting the sensitive attribute. 
\textit{AdvRev} \cite{zhang2018mitigating} de-biases the model by maximizing the loss of predicting the sensitive attribute with loss reversal and gradient projection.
DomainIndep \cite{wang2020towards} trains multiple classifiers, one classifier for each group to explicitly encode separate group information. 
\textit{OBD} \cite{lecun1989optimal} is the pruning-based method,
which uses all training data to compute the saliency without fairness constraints.

\begin{table}[!t]
	\centering
	\caption{Results of accuracy and fairness of different methods on Fitzpatrick-17k dataset, using skin tone as the sensitive attribute. The dark skin is the privileged group with higher accuracy by vanilla training. (pr is the pruning ratio).}
	\label{tab:fitzpatrick}
	\setlength\tabcolsep{5.0pt}
	\renewcommand{\arraystretch}{0.8}
	\resizebox{\columnwidth}{!}{
		\begin{tabular}{lccccccc}
			\toprule
			& & \multicolumn{3}{c}{Accuracy}  & \multicolumn{3}{c}{Fairness}                    \\ \cmidrule(lr){3-5}\cmidrule(lr){6-8}
			Method &  Skin Tone     & Precision & Recall & F1-score & Eopp0 ($\times10^{-3}$) $\downarrow$  & Eopp1 $\downarrow$    & Eodd $\downarrow$    \\ \midrule
			\multirow{4}{*}{Vanilla} & Dark & 0.563          & 0.581     & 0.546          & \multirow{4}{*}{1.331}          & \multirow{4}{*}{0.361}         & \multirow{4}{*}{0.182}          \\
			 & Light & 0.482 & 0.495 & 0.473 &           &          &          \\
			 & Avg. $\uparrow$ & 0.523 & 0.538 & 0.510 &           &          &          \\
			 & Diff. $\downarrow$ & 0.081 & 0.086 & 0.073 &           &          &          \\
			 \midrule
			 \multirow{4}{*}{AdvConf \cite{zhang2018mitigating}} & Dark & 0.506 & 0.562 & 0.506 & \multirow{4}{*}{1.106}          & \multirow{4}{*}{0.339}         & \multirow{4}{*}{0.169}          \\
			 & Light & 0.427 & 0.464 &  0.426        &           &          &\\
			 & Avg. $\uparrow$ & 0.467 & 0.513 & 0.466 &           &          &          \\
			 & Diff. $\downarrow$ & 0.079 & 0.098 & 0.080 &           &          & \\
			 \midrule
			 \multirow{4}{*}{AdvRev \cite{tzeng2015simultaneous}} & Dark & 0.514 & 0.545 & 0.503 & \multirow{4}{*}{1.127}          & \multirow{4}{*}{0.334}         & \multirow{4}{*}{0.166}          \\
			 & Light & 0.489 & 0.469 &  0.457 &           &          &          \\ 
			 & Avg. $\uparrow$ & 0.502 & 0.507 & 0.480 &           &          &          \\
			 & Diff. $\downarrow$ & 0.025 & 0.076 & 0.046 &           &          & \\
			 \midrule
			 
			 \multirow{4}{*}{DomainIndep \cite{wang2020towards}} & Dark & 0.547 & 0.567 & 0.532 & \multirow{4}{*}{1.210}          & \multirow{4}{*}{0.344}         & \multirow{4}{*}{0.172}          \\
			 & Light & 0.455 & 0.480 &  0.451 &           &          &          \\ 
			 & Avg. $\uparrow$ & 0.501 & 0.523 & 0.492 &           &          &          \\
			 & Diff. $\downarrow$ & 0.025 & 0.076 & 0.046 &           &          & \\
			 \midrule
			\multirow{2}{*}{OBD \cite{lecun1989optimal}} & Dark & 0.557 & 0.570 & 0.536 & \multirow{4}{*}{1.244}          & \multirow{4}{*}{0.360}         & \multirow{4}{*}{0.180}          \\
			 & Light & 0.488 & 0.494 & 0.475&           &          &          \\ 
			 (pr=35$\%$) & Avg. $\uparrow$ & 0.523 & 0.532 & 0.506 &           &          &          \\
			 & Diff. $\downarrow$ & 0.069 & 0.076 & 0.061 &           &          & \\
			 \midrule
            \multirow{2}{*}{FairPrune} & Dark & 0.567 & 0.519 & 0.507 & \multirow{4}{*}{\textbf{0.846}} & \multirow{4}{*}{\textbf{0.330}} & \multirow{4}{*}{\textbf{0.165}} \\ 
                & Light & 0.496 & 0.477 & 0.459 & & &   \\
            (pr=35$\%$, $\beta$=0.33) & Avg. $\uparrow$ & 0.531 & 0.498 & 0.483 &           &          &          \\
			 & Diff. $\downarrow$ & 0.071 & \textbf{0.042} & 0.048 &           &          & \\
            \bottomrule
	\end{tabular}}
	\label{table:fitzpatrik}
\end{table}

\noindent
\textbf{Results on Fitzpatrick-17k Dataset.} Table \ref{table:fitzpatrik} shows the accuracy and fairness results of all methods. For accuracy metrics, we report the precision, recall, and F1-score.
From the table, we can see that when achieving a similar level of fairness, FairPrune shows better mean accuracy and a smaller accuracy difference in terms of F1-score compared with two adversarial training baselines \textit{AdvConf} and \textit{AdvRev}.
This is because Fairprune can almost keep the accuracy of the unprivileged group unchanged when balancing the accuracy of two groups.
We can also observe that although vanilla OBD has better overall accuracy than FairPrune, it does not show significant fairness improvements.

\noindent
\textbf{Results on ISIC 2019 Dataset.}
Table \ref{table:isic} shows the comparison of accuracy and fairness results of all methods.
FairPrune achieves significantly better fairness while preserving accuracy.
First, our FairPrune method achieves significantly better fairness than the baselines.
Our FairPrune method achieves $21.0\times 10^{-3}$ Eopp1 and $28.8\times 10^{-3}$ Eodd, which are 57.7\% and 48.3\% lower than $49.7\times 10^{-3}$ and $55.8\times 10^{-3}$ achieved by the best baseline.
Second, the accuracy of FairPrune is comparable to the Vanilla method without fairness constraints, while the F1-score difference between two groups of our method is only 0.003, an order of magnitude lower than the Vanilla training and other methods.

\begin{table}[!t]
	\centering
	\caption{Results of accuracy and fairness of different methods on ISIC 2019 dataset, using gender as the sensitive attribute. The female group is the privileged group with higher accuracy by vanilla training. (pr is the pruning ratio).
	}
	\setlength\tabcolsep{3.0pt}
	\renewcommand{\arraystretch}{0.8}
	\resizebox{\columnwidth}{!}{
		\begin{tabular}{lccccccc}
			\toprule
			& & \multicolumn{3}{c}{Accuracy}  & \multicolumn{3}{c}{Fairness}                    \\ \cmidrule(lr){3-5}\cmidrule(lr){6-8}
			Method &  Gender     & Precision & Recall & F1-score & Eopp0($\times 10^{-3}$)$\downarrow$  & Eopp1($\times 10^{-3}$)$\downarrow$      & Eodd($\times 10^{-3}$)$\downarrow$         \\ \midrule
			\multirow{4}{*}{Vanilla} & Female  &  0.758 & 0.733 & 0.744 & \multirow{4}{*}{6.1} & \multirow{4}{*}{49.7} & \multirow{4}{*}{55.8}         \\
			 & Male & 0.766 & 0.684 & 0.716       \\ 
			 & Avg $\uparrow$    & 0.762 & 0.709 & 0.730 \\ 
			 & Diff $\downarrow$ & 0.008 & 0.049 & 0.028 \\ \midrule
			 \multirow{4}{*}{AdvConf \cite{zhang2018mitigating}} & Female & 0.691 & 0.688 & 0.686  & \multirow{4}{*}{\textbf{4.0}} & \multirow{4}{*}{75.1} & \multirow{4}{*}{79.1}         \\
			 & Male &  0.681 & 0.656 & 0.665           \\ 
			 & Avg $\uparrow$    & 0.686 & 0.672 & 0.675 \\ 
			 & Diff $\downarrow$ & 0.010 & 0.032 & 0.021 \\\midrule
			 \multirow{4}{*}{AdvRev \cite{tzeng2015simultaneous}} & Female & 0.638 & 0.714 & 0.670  & \multirow{4}{*}{5.0} & \multirow{4}{*}{59.2} & \multirow{4}{*}{64.2}         \\
			 & Male &  0.642 & 0.666 & 0.650          \\
			 & Avg $\uparrow$    & 0.640 & 0.690 & 0.660 \\ 
			 & Diff $\downarrow$ & 0.004 & 0.048 & 0.020 \\\midrule
			 \multirow{4}{*}{DomainIndep \cite{wang2020towards}} & Female & 0.782 & 0.693 & 0.729 & \multirow{4}{*}{5.0} & \multirow{4}{*}{74.7} & \multirow{4}{*}{79.7}         \\
			 & Male & 0.783 & 0.653 & 0.697            \\ 
			 & Avg $\uparrow$    & 0.782 & 0.673 & 0.713 \\ 
			 & Diff $\downarrow$ & \textbf{0.001} & 0.040 & 0.032 \\\midrule
            \multirow{2}{*}{OBD \cite{lecun1989optimal}} & Female & 0.771 & 0.734 & 0.749 & \multirow{4}{*}{6.1} & \multirow{4}{*}{55.5} & \multirow{4}{*}{61.6}          \\
			 & Male & 0.762 & 0.678 & 0.711            \\
			 (pr=50$\%$) & Avg $\uparrow$    & 0.767 & 0.706 & 0.730 \\ 
			 & Diff $\downarrow$ & 0.009 & 0.056 & 0.038 \\\midrule
            \multirow{2}{*}{FairPrune} & Female & 0.754 & 0.674 & 0.707  & \multirow{4}{*}{7.8} & \multirow{4}{*}{\textbf{21.0}} & \multirow{4}{*}{\textbf{28.8}}         \\
			 & Male &  0.762 & 0.675 & 0.710 &        \\
			 (pr=50$\%$, $\beta$=0.2)& Avg $\uparrow$    & 0.758 & 0.675 & 0.709 \\ 
			 & Diff $\downarrow$ & 0.008 & \textbf{0.001} & \textbf{0.003} \\\midrule
	\end{tabular}}
	\label{table:isic}
\end{table}

\noindent
\textbf{Ablation Study}. In this section, we discuss how the hyper-parameter $\beta$ and pruning rate will affect the performance of FairPrune.
We apply FairPrune to the model pre-trained on the Fitzpatrick-17k dataset and change $\beta$ defined in Eq.(\ref{equ:combined_score}) to see how the accuracy and fairness performance change.
After that, we fix $\beta$ and change the pruning ratio from 0 to 80\% and repeat the evaluation.

\begin{figure}[t]
	\centering
	\includegraphics[width=1.0\columnwidth]{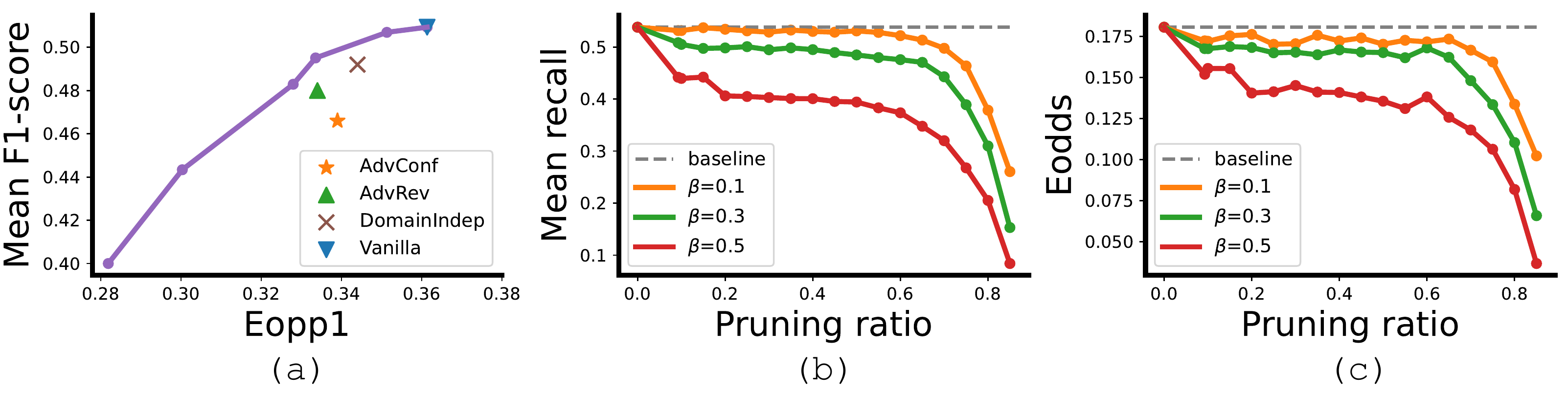}
	\caption{Ablation study on hyper-parameter $\beta$ and pruning ratio for skin tone fairness on Fitzpatrick-17k dataset.}
	\label{fig:ablation study}
\end{figure}

In Fig. \ref{fig:ablation study}(a), we show the accuracy (F1-Score) and fairness (Eopp1) trade-off by varying $\beta$ under a fixed pruning ratio of 35\%.
Points on the upper left of this figure represent better fairness (smaller Eopp1) and higher accuracy (larger F1-score). 
With varying $\beta$, our methods achieve consistent better fairness and accuracy trade-off.
We can see that when achieving the same fairness, FairPrune has higher accuracy than the baselines. 
At the same accuracy, FairPrune can also show better fairness than the baselines.
Therefore, by tuning $\beta$, the trade-off between fairness and accuracy can be adjusted to satisfy the user requirement.

Fig. \ref{fig:ablation study}(b) and Fig. \ref{fig:ablation study}(c) show the effect of the pruning ratio on accuracy and fairness, respectively.
It can be seen that the mean recall drops slowly and then quickly, as well as the fairness metrics Eopp1.
Although the fairness improvement can be larger at a larger pruning ratio (e.g., $pruning\ ratio=0.8$), the accuracy may also be lower.
Therefore, we suggest that the optimal pruning ratio usually exists at a lower pruning ratio where we do not see a significant accuracy drop on both groups but better fairness.

\section{Conclusion}
In this paper, we propose FairPrune, 
a method to achieve fairness by pruning. 
Based on our observation that each model parameter has different importance for different groups' accuracy,
by pruning the parameters based on this importance difference, 
we can reduce the accuracy difference between the privileged group and the unprivileged group to improve fairness without a large accuracy drop.
We evaluate our method on two skin lesion analysis datasets over multiple sensitive attributes. 
The experiment results show that 
FairPrune can greatly improve fairness while keeping the average accuracy of both groups as high as possible.

%
%
%
\bibliographystyle{splncs04}
\bibliography{reference.bib}

\end{document}